\documentclass[%
 aps,
 prb,
 twocolumn,
 reprint,%
citeautoscript,
showkeys
]{revtex4-1}

\usepackage{chemformula}
\usepackage{graphicx}%
\usepackage{dcolumn}%
\usepackage{siunitx}
\usepackage{bm}%
\usepackage{placeins}
\begin{document}

\title[]{Transferability in Machine Learning for Electronic Structure\\ via the Molecular Orbital Basis} %

\author{Matthew Welborn}
\author{Lixue Cheng}%

\author{Thomas F. Miller III}
 \email{tfm@caltech.edu.}
\affiliation{%
Division of Chemistry and Chemical Engineering, California Institute of Technology, Pasadena, CA 91125, USA%
}%

\date{\today}%

\begin{abstract}
We present a machine learning (ML) method for  predicting   electronic structure correlation energies using Hartree-Fock input.  
The total correlation energy is expressed in terms of individual and pair contributions from occupied molecular orbitals, and 
Gaussian process regression is used to predict these %
contributions from a feature set that is based on molecular orbital properties, such as Fock, Coulomb, and exchange matrix elements.
With the aim of  maximizing transferability across chemical systems and compactness of the feature set, we avoid the usual specification of ML features in terms of atom- or geometry-specific information, such atom/element-types, bond-types, or local molecular structure. 
ML predictions of MP2 and CCSD energies are presented for a range of systems, demonstrating that the method maintains accuracy while providing transferability both within and across chemical families; this includes   predictions for molecules with atom-types and elements that are not included in the training set. 
The method 
holds promise both in its current form and as a proof-of-principle for the use of ML in the  design of  %
generalized density-matrix functionals. 
\end{abstract}

\maketitle

\section{Introduction}
Recent interest in the use of machine learning (ML)  for electronic structure has  focused on models that are formulated in terms of atom- and geometry-specific features, such as atom-types and bonding connectivities.
The advantage of this approach is that it can yield excellent accuracy with computational cost that is comparable to classical force fields.\cite{Bartok2010,rupp2012fast,VonLilienfeld2013,hansen2013assessment, Ceriotti2014,ramakrishnan2015big,Tuckerman,Behler2016,kearnes2016molecular,Paesani2016, schutt2017quantum,Smith2017,wu2018moleculenet,Nguyen2018,Yao2018, Fujikake2018}  %
However, a %
disadvantage of this approach is that building a ML model to describe a diverse %
set of elements and  chemistries requires training with respect to a number of features that grows quickly with the number of atom- and bond-types, %
and also requires vast amounts of reference data for %
the selection and training of those features; these issues have hindered the degree of 
chemical transferability of existing ML models for electronic structure.
For example, previous methods have not demonstrated predictions for molecules with chemical elements that are not included in the training data.

In this work, we focus on the more modest goal of using ML to describe the post-Hartree-Fock correlation energy.  Assuming willingness to incur the cost of a Hartree-Fock self-consistent field (SCF) calculation, we aim to describe the correlation energy associated with perturbation theory,\cite{Moller1934} coupled-cluster theory,\cite{Cizek1966} or other post-Hartree-Fock methods.
Our approach focuses on training not %
with respect to atom-based features, but instead using features based on the Hartree-Fock molecular orbitals (MOs), which have no explicit dependence on %
 the underlying atom-types %
and  may thus be expected to provide greater chemical transferability.

For a general post-Hartree-Fock electronic structure method, the correlation energy may be expressed via Nesbet's theorem as a sum over occupied MOs\cite{Nesbet1958}  %
\begin{equation}
\label{Ecorr}
E_\textrm{c} 
=\sum^{\textrm{occ}}_{ij}\epsilon_{ij}.
\end{equation}
Our strategy %
is to use ML to describe the diagonal and off-diagonal contributions to this sum,
\begin{equation}
\label{energies}
\epsilon_{ii} = \epsilon_\textrm{d}\left(\textbf{f}_{i} \right) \quad\textrm{and}\quad
\epsilon_{ij} = \epsilon_\textrm{o}\left(\textbf{f}_{ij} \right),
\end{equation}
respectively, where $\textbf{f}_{i}$ is a vector of features associated with the $i^{\textrm{th}}$ occupied MO, and $\textbf{f}_{ij}$ is a vector of features associated with the $i,j$ pair of occupied MOs.  %
Employing this strategy in the representation of localized MOs (LMOs), for which Eq.~\ref{Ecorr} also holds,  %
leads to a ML model 
that is compact with respect to the number of  features and that is both chemically accurate and encouragingly transferable across chemical systems. 

\section{Feature Design and Selection}\label{sec:feature_design}
 All ML features used  in this study are elements of  %
the Fock matrix $\mathbf{F}$, Coulomb  matrix $\mathbf{J}$, or  exchange matrix $\mathbf{K}$. 
 With the aim of maximizing transferability of the features, we represent the matrices in the LMO basis. %
Only matrix elements associated with the subset of valence occupied and virtual LMOs are included as ML features; occupied core orbitals are excluded, as the post-Hartree-Fock calculations employ the frozen core approximation, and 
 the valence virtual orbitals are defined by projection onto a minimal basis (details in Sec.~\ref{sec:calc-details}).\cite{Subotnik2005} %

For a given $i,j$ pair of occupied LMOs, the total feature vector $\mathbf{f}_{ij}$ is comprised of feature vectors associated with elements of the Fock, Coulomb, and exchange matrices,
\begin{align}
\label{totvector}
\mathbf{f}_{ij} & = \left(\mathbf{f}_{ij}^{(\mathbf{F})}, \mathbf{f}_{ij}^{(\mathbf{J})},\mathbf{f}_{ij}^{(\mathbf{K})}\right).
\end{align} 
These composite vectors involve matrix elements  from the occupied-occupied, occupied-virtual, and virtual-virtual blocks of the matrices, such that 
\begin{align}
\label{Eq:ODfeatures}
\mathbf{f}_{ij}^{(\mathbf{F})} & = \left(F_{ii}, F_{ij}, F_{jj},\mathbf{F}^\textrm{vv}_{ij}\right) \\ 
\mathbf{f}_{ij}^{(\mathbf{J})} & = \left(J_{ii}, J_{ij}, J_{jj}, \mathbf{J}^\textrm{v}_{i}, \mathbf{J}^\textrm{v}_{j},\mathbf{J}^\textrm{vv}_{ij}\right) \nonumber\\
\mathbf{f}_{ij}^{(\mathbf{K})} & = \left(K_{ij}, \mathbf{K}^\textrm{v}_{i}, \mathbf{K}^\textrm{v}_{j}
,\mathbf{K}^\textrm{vv}_{ij}\right), \nonumber
\end{align}
 where 
 terms are sorted %
 with respect to $i$ and $j$ such that $F_{ii}<F_{jj}$. 
 This sorting guarantees that $\epsilon_{ij}$ = $\epsilon_{ji}$.
 The vectors
$\mathbf{J}^\textrm{v}_{i}$, $\mathbf{J}^\textrm{v}_{j}$, $\mathbf{K}^\textrm{v}_{i}$, and $\mathbf{K}^\textrm{v}_{j}$ %
include matrix elements associated with localized valence virtual orbitals (indexed $a,b,c,\ldots$) such that
\begin{align}
\label{EqODfeatures2}
\mathbf{J}^\textrm{v}_{i}  & = \left(J_{ia},J_{ib},J_{ic}, ... \right)\\
\mathbf{K}^\textrm{v}_{i}  & = \left(K_{ia},K_{ib},K_{ic}, ...\right) \nonumber
\end{align}
and likewise for $\mathbf{J}^\textrm{v}_{j}$ and $\mathbf{K}^\textrm{v}_{j}$.
The localized valence virtual orbitals associated with the matrix elements in $\mathbf{J}^\textrm{v}_{i}$
and $\mathbf{K}^\textrm{v}_{i}$ are selected and sorted on the basis of having the largest off-diagonal Coulomb matrix elements, %
such that $J_{ia}>J_{ib}>J_{ic}$, etc.; likewise for 
$\mathbf{J}^\textrm{v}_{j}$
and $\mathbf{K}^\textrm{v}_{j}$.
Note that %
the valence virtual LMO associated with $J_{ia}$ 
is the same as for $K_{ia}$, but it 
need not be the same as that associated with $J_{ja}$. %
Finally, the matrices 
$\mathbf{F}^\textrm{vv}_{ij}, \mathbf{J}^\textrm{vv}_{ij},$ and $\mathbf{K}^\textrm{vv}_{ij}$ in Eq.~\ref{Eq:ODfeatures} contain  virtual-virtual matrix elements corresponding to localized valence virtual orbitals %
that are selected and sorted such that %
$J_{ia}+J_{ja}>J_{ib}+J_{jb}$, etc.; only the upper diagonal of these matrices comprise independent features and are included. 
Because they appear in the cluster amplitude equations of MP2 and CCSD, virtual-virtual matrix elements are potentially %
informative features for the prediction of pair correlation energies.

Appropriate sorting of the virtual LMOs was found to be important for achieving transferability as the GP regression is sensitive to permutations of elements within the feature vector. 
$J_{ia}$ acts as a proxy for spatial distance.
As dynamical electron correlation is ``near-sighted,''\cite{Boughton1993} the spatially closest valence virtual LMOs are also likely to be most important to the pair correlation energy. 
In a large system, the number of included valence virtual LMOs must be limited, %
and sorting ensures that these most important elements are included in the feature vector. 
Any distance-based cutoff procedure is subject to discontinuities in the energy if valence virtual LMOs move in or out of the cutoff region. As in local correlation methods, sufficiently large cutoffs must be chosen to ensure that the energy surface is acceptably smooth.\cite{Russ2004} The near-sighted nature of dynamical correlation also leads to the expectation that the number of needed features based on valence virtual LMOs quickly saturates with system size, which is confirmed in the results presented below. 

The resulting features are invariant with respect to rotation and translation of the system, invariant to rotations among the occupied MOs that precede localization, smooth with respect to molecular geometry, and unique for each geometry -- to the extent that the employed orbital localization method has these properties.
In this work, we employ the Intrinsic Bond Orbital method which has been shown to yield unique LMOs which vary smoothly with geometry.\cite{Knizia2013IBO,Knizia2015} 
By construction, the features yield a model with sufficient flexibility to describe dissociation into two closed-shell fragments. 
In the dissociated limit, features corresponding to occupied pairs with both $i$ and $j$ on one fragment contain no information about the other fragment. 
For occupied pairs $i$ and $j$ that span fragments, 
by including dissociated fragments in the training data,
the ML model is trained to predict that 
$\epsilon_{ij}$ vanishes as features involving both $i$ and $j$ (e.g. $J_{ij}$) go to zero.

For each occupied LMO used to describe the diagonal contributions to the correlation energy, $\epsilon_{\textrm{d}}(\mathbf{f}_{i})$ in Eq.~\ref{energies}, the total feature vector $\mathbf{f}_{i}$ is obtained by keeping only the unique  terms in $\mathbf{f}_{ii}$.

\section{Calculation %
Details}
\label{sec:calc-details}
All Hartree-Fock, second-order M{\o}ller-Plessett perturbation theory (MP2),\cite{Moller1934} and coupled-cluster with singles and doubles (CCSD)\cite{Cizek1966} calculations are performed using the \textsc{Molpro} 2018.0 software package.\cite{Werner2012} 
Unless otherwise stated, calculations employ the cc-pVTZ basis set.\cite{Dunning1989}
The frozen-core approximation is employed for correlated calculations. 

Valence occupied and virtual LMOs are generated using the Intrinsic Bond Orbital method\cite{Knizia2013IBO} with a localization threshold of $10^{-12}$;  core orbitals are excluded from localization. %
This method is detailed in Ref. \citenum{Knizia2013IBO} and summarized here. 
A set of Intrinsic Atomic Orbitals (IAOs) is formed by polarizing a minimal basis of free-atom atomic orbitals to form a set of the same size that can exactly represent the occupied MOs of a given Slater determinant. 
The IAOs are then partitioned into an occupied subset, whose span is the occupied MOs, and a virtual subset, whose span defines the valence virtual MOs. 
These two sets are localized using the Pipek-Mezey criterion\cite{Pipek1989} to form the occupied and valence virtual LMOs. The subset of valence virtual MOs are readily localized.\cite{Lu2004,Subotnik2005,Hoyvik2012} 

For the selected features, Gaussian process regression (GPR)\cite{rasmussen2006} of $\epsilon_\textrm{d}$ and $\epsilon_\textrm{o}$ in Eq.~\ref{energies} is separately performed with the \textsc{GPy} software package. \cite{gpy2014}
The Mat\'ern 5/2 kernel\cite{rasmussen2006} is employed with white-noise regularization. A single length scale is used for all features, resulting in a total of three kernel hyperparameters. %
The scaled conjugate gradient method\cite{moller1993scaled} is used to minimize the negative log marginal likelihood objective with respect to the kernel hyperparamters. Kernel ridge regression\cite{KRR} was also explored but was not found to lead to more accurate predictions than GPR.

In all cases, training and test geometries are generated from an \textit{ab initio} molecular dynamics trajectory performed with the \textsc{Q-Chem} 5.0 software package,\cite{Shao2015} 
using the B3LYP\cite{Vosko1980,Lee1988,Becke1993,Stephens1994}/6-31g*\cite{Hariharan1973} level of theory and a Langevin thermostat\cite{Bussi2007} at 350 K.
Geometries are sampled from the trajectories at 50 fs intervals. 
For each training geometry, data associated with all occupied orbitals is employed for training, although results are unchanged if a consistent number of orbital pairs is randomly selected from training geometries.

To avoid overfitting, the total number of features should be reduced prior to training. We prioritize features based on the intuition that features involving two occupied LMOs (e.g. $J_{ij}$) are more important than features involving one occupied and one valence virtual LMO (e.g. $J_{ia}$), which are in turn more important than features involving two valence virtual LMOs (e.g. $J_{aa}$). %
This intuition largely agrees with feature Gini importance rankings determined automatically via Decision Tree Regression (DTR),\cite{DTR} while avoiding pathologies found using naive application of the latter for some cases. %
Such pathologies can arise from the fact that DTR Gini importance ranks features by how well they lead to separate clusters in feature space, with less regard for variability within those clusters.\cite{DTR} 
Optimal features for ML in our application should describe  variability both within and between these clusters. 
This leads to problems for the DTR method in cases such as alkanes that have only one type of occupied LMO (i.e., sigma bonds) and thus yield no distinct clusters; in these cases, naive application of DTR fails to select any features. Nonetheless, we acknowledge that more sophisticated automatic feature selection methods are available and will be investigated in future work.
For the purposes of this work, we monitor potential overfitting using out-of-sample testing; during training, we hold out a subset of the training set and confirm that the errors from this subset are similar to those from the training set.
Employed features sets used in this study are listed in Tab. \ref{table:feature_sets}.

\begin{table}[htbp]
\caption{Employed feature sets, %
and the number of features for the diagonal (\#$\mathbf{f}_i$) and off-diagonal (\#$\mathbf{f}_{ij}$) pairs.}
\begin{ruledtabular}
\begin{tabular}{lp{2.3in}rr}
Set&Description&\#$\mathbf{f}_i$&\#$\mathbf{f}_{ij}$\\ \hline
A & Features corresponding to the occupied-occupied and occupied-virtual blocks of \textbf{F}, \textbf{J}, and \textbf{K}, including only  the first four localized valence virtual orbitals. %
& 10 & 23\\ 

B & Feature Set A, with $F_{aa}$, $J_{aa}$, and $K_{ab}$ also included in $\mathbf{f}_{i}$. & 13 & 23\\

C & $\mathbf{f}_i = \left(F_{ii}, F_{aa}, J_{ii}, J_{ia}, J_{aa}, K_{ia}\right)$ & 6 & 7\\
  & $\mathbf{f}_{ij} = \left(F_{ii}, F_{ab}, J_{ii}, J_{ij}, J_{jj}, K_{ij}, K_{ja}\right)$\\
\end{tabular}
\end{ruledtabular}
\label{table:feature_sets}
\end{table}

\section{Results} %

\subsection{Transferability among geometries}
\begin{figure*}[htbp]
\includegraphics[width=1.0\textwidth]{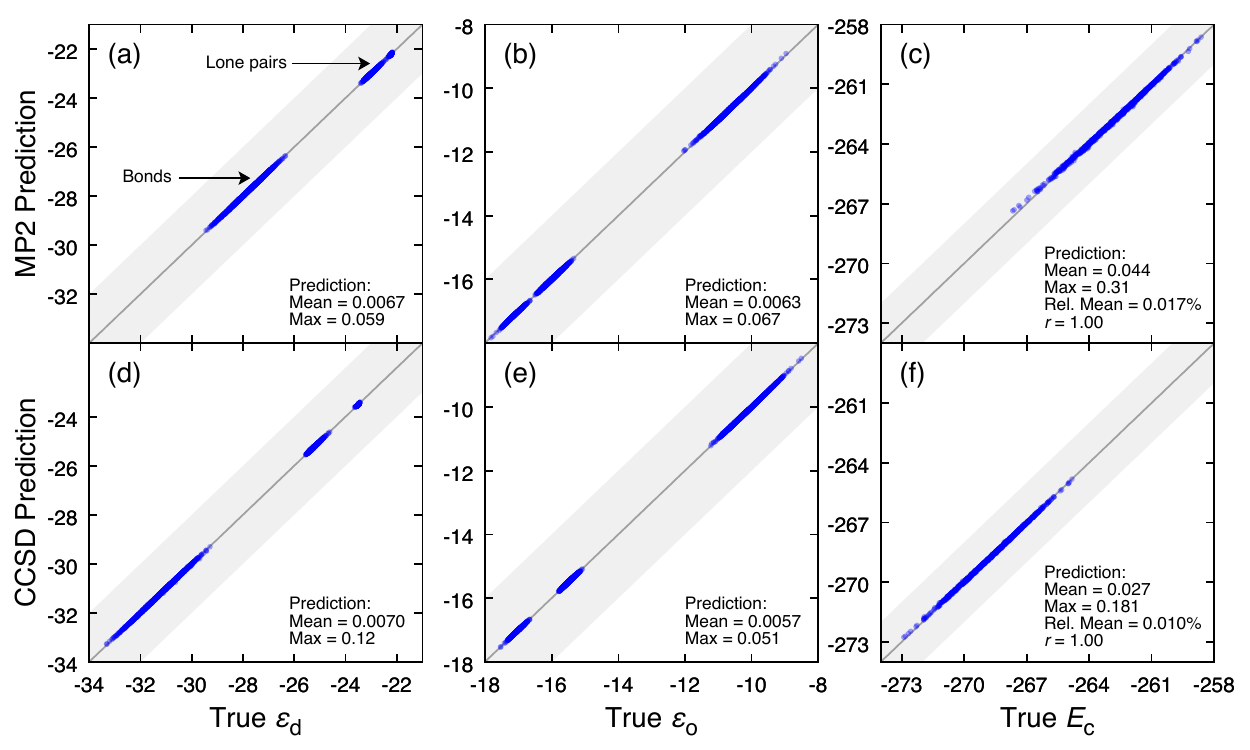}
\caption{
ML predictions of MP2 (a-c) and CCSD (d-f) results for a water molecule, training on  200  geometries and predicting for 1000 geometries, 
including $\epsilon_{\textrm{d}}$ (a,d) and $\epsilon_{\textrm{o}}$ (b,e) for the pairs of occupied orbitals, as well as the total correlation energies (c,f).  Mean absolute errors (Mean), maximum absolute errors (Max), Mean Errors as a fraction of total correlation energy (Rel. Mean), and the Pearson correlation coefficient ($r$) are reported; all energies in mH. The guideline indicates zero error, with the region of %
up to 2 mH error indicated via shading. %
}
\label{figure:water-self-prediction}
\end{figure*}

For the example of a single water molecule, we begin by training the ML model on a subset of geometries to predict the correlation energy at other geometries. For both the MP2 and CCSD levels of theory, the diagonal ($\epsilon_{\textrm{d}}$) and off-diagonal ($\epsilon_{\textrm{o}}$) contributions to the correlation energy are separately trained using Feature Set A (Tab.~\ref{table:feature_sets}) with 200  geometries, and the  resulting ML predictions for a superset of 1000 geometries are presented in Fig.~\ref{figure:water-self-prediction}.
Errors are summarized in terms of mean absolute error (Mean Error), maximum absolute error (Max Error), and Mean Error as a percentage of the mean total correlation energy (Rel. Mean Error); 
energies are reported in milliHartrees (mH) throughout the paper. 
The Pearson correlation coefficient ($r$) is also reported as a measure of correlation between the ML predictions and the true values;\cite{Pearson1896} a value of $r=1$ indicates perfect correlation, $r=0$ indicates no correlation, and $r=-1$ indicates perfect anticorrelation. Note that a value of $r=1$ does not imply that the slope of the relationship is unity.

As illustrated for the diagonal contributions in Fig.~\ref{figure:water-self-prediction}a, the individual contributions to the correlation energy exhibit clusters associated with common physical origins (i.e., $\sigma$-bonding vs.~lone-pair orbitals). 
 For both the diagonal and off-diagonal contributions, the agreement between the ML prediction and the reference result is excellent, %
 leading to predictions for the total correlation energy that are well within chemical accuracy. %
For all examples studied in this work, we find the quality of ML predictions for MP2 and CCSD to be qualitatively similar (as in Fig. 1); MP2 results are thus presented in the SI for the remainder. %

Table \ref{table:self-prediction}  summarizes the corresponding results for %
other small molecules, with $\epsilon_{\textrm{d}}$ and $\epsilon_{\textrm{o}}$ trained on a subset of geometries and used to predict the CCSD correlation energy for other geometries.  
The molecules %
range in size from \ch{H2} to benzene.
Feature Set A is used in all cases, except for ethane, for which Feature Set B was needed to achieve comparable accuracy. The number of geometries included in the training set and testing superset are indicated in the table. 
In general, the Mean Error for the correlation energy is much less than 1 mH, and the Max Error is also in the range of chemical accuracy. 
Note that we are predicting the correlation energy for these molecules with a Rel. Mean Error that is  0.1\% or less for all cases.

Table~\ref{table:self-prediction} also illustrates the sensitivity of the ML predictions to changing the number of geometries in the training set (for ethane, formic acid, and difluoromethane) or the employed basis set (for water). 
Although the additional geometries for these cases lead to better ML prediction accuracy, further improvement with additional geometries eventually becomes limited by the baseline self-training error of the employed GPR method. %
The water results for basis sets ranging from double-zeta to quintuple-zeta make clear that the ML prediction is not sensitive to the employed basis set.
\begin{table}[htbp]
\caption{
ML predictions of CCSD correlation energies for a collection of small molecules, with the number of training and testing geometries indicated. A more detailed breakdown of the diagonal and off-diagonal contributions to the correlation energy errors is presented in Tab.~S1. 
} 
\begin{ruledtabular}
\begin{tabular}{lrrccccccc}
&\multicolumn{2}{c}{Geometries}&\multicolumn{2}{c}{Error (mH)}&\multicolumn{2}{c}{Rel. Error(\%)}\\
Molecule&Train&Test&Mean&Max&Mean&Max\\ \hline
\ch{H2}	&	50	&	100	&	0.00	&	0.00	&	0.00	&	0.01	\\
\ch{N2}	&	50	&	100	&	0.06	&	0.19	&	0.01	&	0.05	\\
\ch{F2}	&	50	&	100	&	0.03	&	0.18	&	0.00	&	0.03	\\
\ch{HF}	&	50	&	100	&	0.03	&	0.23	&	0.01	&	0.08	\\
\ch{NH3}	&	50	&	100	&	0.16	&	0.57	&	0.06	&	0.23	\\
\ch{CH4}	&	50	&	100	&	0.03	&	0.10	&	0.01	&	0.05	\\
\ch{CO}	&	50	&	100	&	0.03	&	0.07	&	0.01	&	0.02	\\
\ch{CO2}	&	50	&	100	&	0.04	&	0.17	&	0.01	&	0.03	\\
\ch{HCN}	&	50	&	100	&	0.04	&	0.17	&	0.01	&	0.05	\\
\ch{HNC}	&	50	&	100	&	0.09	&	0.45	&	0.03	&	0.13	\\
\ch{C2H2}	&	50	&	100	&	0.21	&	0.61	&	0.06	&	0.19	\\
\ch{C2H4}	&	50	&	100	&	0.30	&	0.75	&	0.08	&	0.21	\\
\ch{C2H6}$^{\dag}$	&	50	&	1000	&	0.33	&	1.27	&	0.08	&	0.31	\\
	&	200	&	1000	&	0.21	&	1.22	&	0.05	&	0.30	\\
\ch{CH2O}	&	50	&	100	&	0.09	&	0.33	&	0.02	&	0.08	\\
\ch{HCO2H}$^{\dag}$	&	50	&	1000	&	0.40	&	1.24	&	0.06	&	0.19	\\
	&	100	&	1000	&	0.27	&	0.86	&	0.04	&	0.14	\\
\ch{CH3OH}	&	50	&	100	&	0.24	&	0.93	&	0.05	&	0.32	\\
\ch{CH2F2}$^{\dag}$	&	50	&	1000	&	0.73	&	2.94	&	0.11	&	0.43	\\
	&	100	&	1000	&	0.56	&	2.05	&	0.08	&	0.30	\\
\ch{C6H6}	&	50	&	100	&	0.30	&	1.19	&	0.03	&	0.12	\\
													\\
\ch{H2O}$^{\ddag}$													\\
\textit{cc-pVDZ}	&	50	&	200	&	0.05	&	0.22	&	0.02	&	0.10	\\
\textit{cc-pVTZ}	&	50	&	200	&	0.04	&	0.14	&	0.02	&	0.05	\\
\textit{cc-pVQZ}	&	50	&	200	&	0.05	&	0.20	&	0.02	&	0.07	\\
\textit{cc-pV5Z}	&	50	&	200	&	0.08	&	0.37	&	0.03	&	0.13	\\
\end{tabular}
\end{ruledtabular}
$^{\dag}$Two sizes of training sets are presented to illustrate error reduction.
$^{\ddag}$Results for several basis sets provided.

\label{table:self-prediction}
\end{table}

\subsection{Transferability within a molecular family}\label{sec:within_family}

\begin{figure}[!htbp]
\includegraphics[width=1.0\columnwidth]{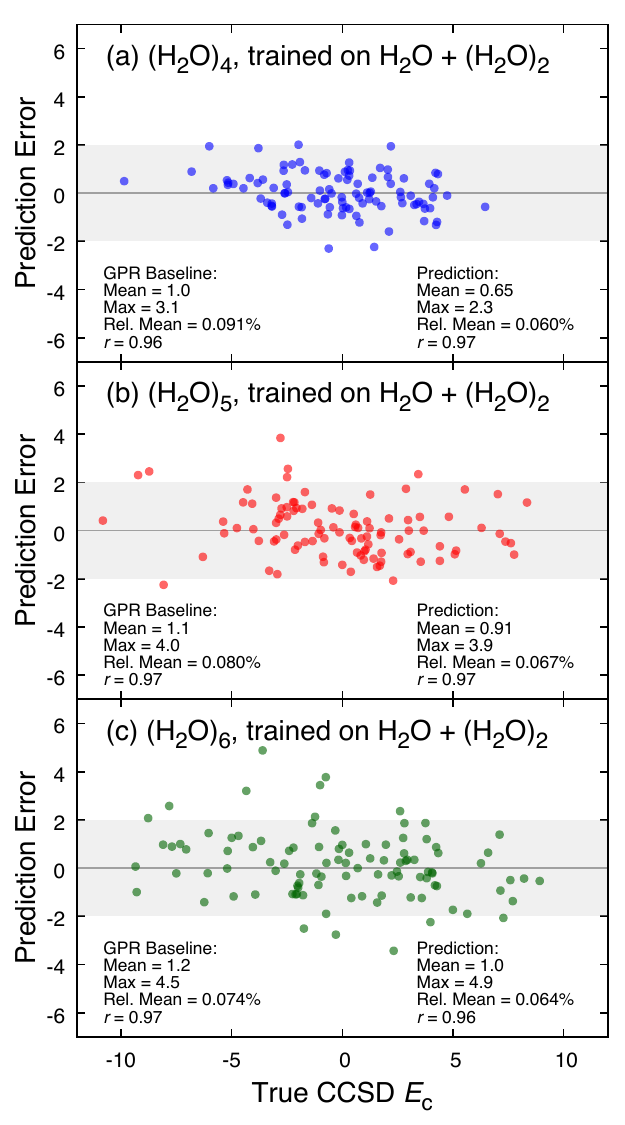}
\caption{
ML predictions of CCSD correlation energies for water tetramer, pentamer, and hexamer, with a ML model obtained from training on the water monomer and dimer. %
ML prediction errors are plotted versus the true CCSD correlation energy. (See SI for corresponding plots of error versus the CCSD total energy.)
Parallelity error is removed via  a 
global shift in the predicted energies of the tetramer, pentamer, and hexamer by 1.7, 2.1, and 3.2 mH, respectively. 
GPR baseline errors correspond to the self-training error of the ML model, providing an expectation for the lowest possible error of the ML model in the employed GPR framework. 
The true CCSD energies are plotted relative to their median.
All energies reported in mH.
}
\label{figure:water_clusters}
\end{figure}

We now explore the degree to which a ML model trained on one molecular system can be used to describe a different system, focusing first on transferability within a molecular family.  Fig.~\ref{figure:water_clusters} shows results for water clusters (tetramer, pentamer, and hexamer) based on training data that includes only the water monomer and dimer. 

The ML model is trained on 200 water monomer and 300 water dimer geometries, and predictions are made for 100 geometries of each of the larger clusters.
To avoid overfitting based on the monomer and dimer input, we employ the smaller Feature Set C.

Figure \ref{figure:water_clusters} shows ML predictions of the CCSD energy of water (a) tetramers, (b) pentamers, and (c) and hexamers. 
In these predictions, the absolute zero of energy is shifted to compare relative energies on the cluster potential energy surface (i.e. parallelity errors are removed); the sizes of these shifts are reported in the caption. 
For all three clusters, the observed Rel. Mean Errors of 0.06-0.07\% are comparable to those reported in Tab. \ref{table:self-prediction}, and the Pearson correlation coefficients exceed 0.95.  %

Although the results in Fig.~\ref{figure:water_clusters} are encouraging in terms of accuracy, additional analysis suggests that more sophisticated regression methods will lead to further improvements.  
To illustrate this, each panel of the figure reports the calculated GPR baseline accuracy, determined via characterizing the self-training error with the employed GPR method.  
For each size of water cluster, a ML model is trained and tested on the same set of 100 geometries; 
this establishes the smallest error that can be expected of the predictions within the current ML framework which maximizes model likelihood rather than minimizing training error. 
The fact that the prediction errors for the ML model for the water clusters are very similar to the GPR baseline error in Fig.~\ref{figure:water_clusters} suggests that the prediction error is dominated by the self-training error of the GPR rather than from a lack of transferability of the ML model trained on water monomers and dimers to larger clusters.
Further refinement of the employed regression method will potentially reduce the baseline error and therefore improve ML predictions.

\begin{figure}[htbp]
\includegraphics[width=1.0\columnwidth]{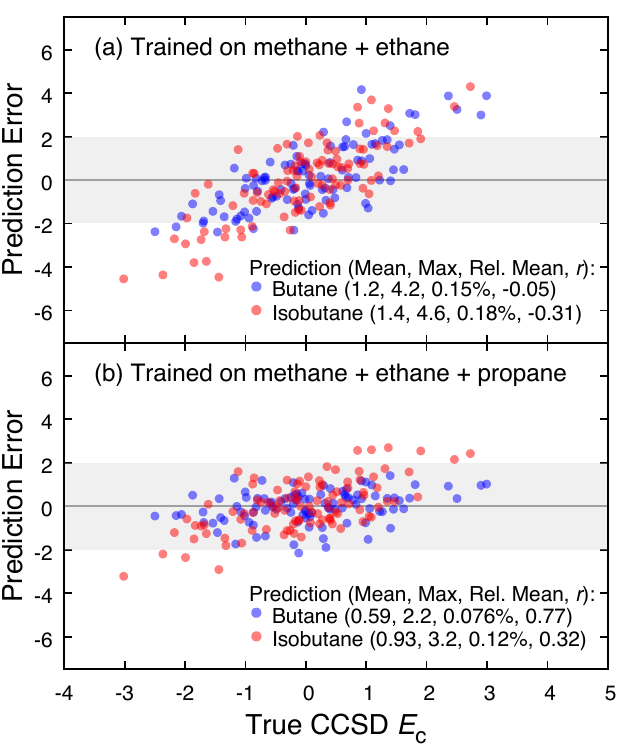}
\caption{
ML predictions of CCSD correlation energies for butane and isobutane, with a ML model obtained from training on (a) methane and ethane and (b) propane in addition. 
ML prediction errors are plotted versus the true CCSD correlation energy.
Parallelity error is removed via a
global shift in the predicted energies of butane and isobutane by (a) 25 and 16 mH and (b) 3.3 and 0.73 mH, respectively. 
The Mean and Max GPR baseline errors for butane are 0.58 and 1.5 mH, respectively. For isobutane, these errors are 0.53 and 1.9 mH.
The GPR baseline Pearson correlation coefficients for butane and isobutane are both 0.79. %
The true CCSD energies are plotted relative to their median. All energies reported in mH.
}
\label{figure:alkanes}
\end{figure}

As a second example, we examine transferability within a family of covalently bonded molecules by  predicting butane and isobutane CCSD energies from shorter alkane training data. 
The ML model is first trained on 100 methane and 300 ethane geometries using Feature Set B, %
and Fig.~\ref{figure:alkanes}a presents the resulting ML predictions for  100 geometries of butane and isobutane. 
Although the Mean Errors are not large (1.2 and 1.4 mH), the Rel. Mean Errors are over twice those obtained for the water cluster series, and
the Mean and Max errors associated with the baseline GPR accuracy (reported in caption) are smaller than the prediction errors. 
Moreover, the correlation coefficients are significantly reduced (-0.05 and -0.31) compared the previous examples, 
although this is partly due to the small range of values for the true CCSD correlation energies in the test set.
These results suggest that additional training data would improve prediction accuracy.

The effect of including additional alkane training data is tested in  Fig.~\ref{figure:alkanes}b, which presents results for which the ML model is retrained with the training data set expanded to include 50 propane geometries.  
The prediction errors and correlation coefficients for butane and isobutane are both substantially 
improved upon inclusion of the propane data, with the butane prediction errors dropping to the GPR baseline while the isobutane prediction errors remain above the GPR baseline. 
Specifically, the correlation coefficients increase to 0.77 and 0.32 for butane and isobutane, respectively, as compared to a GPR baseline correlation coefficient of 0.79 for both molecules. 

Comparison of the ML prediction errors in Figs.~\ref{figure:alkanes}a and \ref{figure:alkanes}b is sensible from the perspective of the carbon atom-types that are included in the training data.  The unbranched butane molecule includes only primary and secondary carbons, whereas isobutane includes a tertiary carbon atom.  In Fig.~\ref{figure:alkanes}a, the training data includes examples of neither secondary nor tertiary carbon atoms; it is thus notable how well the ML model predicts the energies for butane and isobutane, both of which include atom-types that are not included in the training data.  In Fig.~\ref{figure:alkanes}b, the propane training data provides information about secondary carbons to the particular benefit of the butane ML predictions, whereas the isobutane errors, while improved,
remain slightly larger since tertiary carbon examples are still not included in the training data.  Regardless, these results directly illustrate that the ML model exhibits encouraging transferability, 
provides good prediction accuracy even for molecules with atom-types that are not included in the training data,
and demonstrates systematic improvability as the training data increasingly represent chemical environments that appear in the test data.

\subsection{Transferability across molecules and elements}

\begin{figure}[htbp]
\includegraphics[width=1.0\columnwidth]{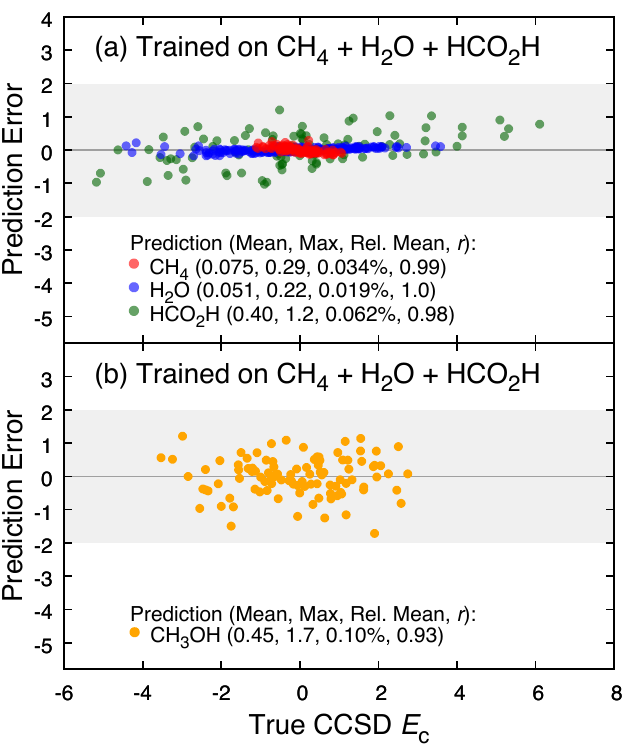}
\caption{
Using a model trained on water, methane, and formic acid, ML predictions of CCSD correlation energy for (a) these same three molecules and (b) methanol. 
ML prediction errors are plotted versus the true CCSD correlation energy.
In panel (b), parallelity error is removed via a global shift in the predicted energy by 3.5 mH.
The true CCSD energies are plotted relative to their median. All energies reported in mH.
}
\label{figure:methanol}
\end{figure}

Figure  \ref{figure:methanol} explores ML predictions for methanol using a  training set that contains methane, water, and formic acid.  For this example, the  training molecules include similar bond-types and the same elements as methanol, but different bonding connectivity.  The ML model is trained on 50 geometries each of methane, water, and formic acid, using Feature Set A; the model is then used to predict CCSD energies  for a superset 
of 100 geometries of each of the molecules in the training set (Fig.~\ref{figure:methanol}a) and for 100 geometries of the methanol molecule (Fig.~\ref{figure:methanol}b).

Fig.~\ref{figure:methanol}a first shows predictions for the molecules that are represented  within the training set.
The resulting errors are similar to those observed when separate models are trained for each of these molecules individually (Tab. \ref{table:self-prediction}), indicating that the ML model has the flexibility to simultaneously describe this group of chemically distinct molecules. 

In Fig. \ref{figure:methanol}b, the same ML model is used to predict the CCSD energy of  methanol, which is not represented in the training set. 
The resulting Mean and Max Errors for methanol are comparable to those for the molecules in the training set, and notably, these errors are only about twice as large as those %
obtained from training methanol on itself (Tab.~\ref{table:self-prediction}). 
Moreover, the Pearson correlation coefficients are high in all cases.
These results demonstrate that the ML model successfully transfers information learned about pair correlation energies in methane, water, and formic acid toward the prediction of methanol, while preserving chemical accuracy.

\begin{figure}[htbp]
\includegraphics[width=1.0\columnwidth]{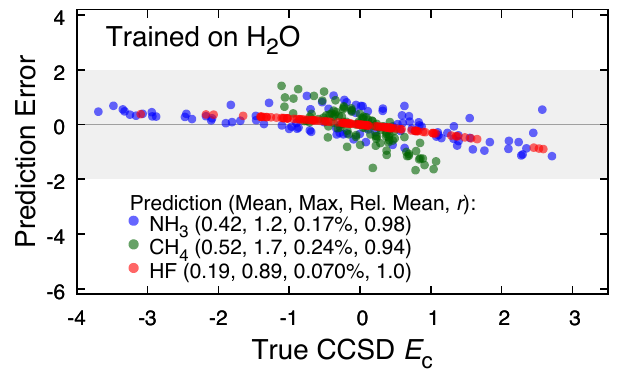}
\caption{
ML predictions of CCSD correlation energies for ammonia, methane, and hydrogen fluoride, with the ML model obtained from training on water. 
ML prediction errors are plotted versus the true CCSD correlation energy.
Parallelity error is removed via a 
global shift in the predicted energies of ammonia, methane, and hydrogen fluoride by 3.4, 16, and 5.6 mH, respectively. 
The true CCSD energies are plotted relative to their median.
All energies are reported in mH.
}
\label{figure:water_to_other}
\end{figure}

Finally, as an extreme test of transferability of the ML model,  we explore cases for which predictions are made on molecules with chemical elements that do not appear in the training set. 
Figure \ref{figure:water_to_other} shows the ML predictions for the CCSD energies of  100 geometries each of ammonia, methane, and hydrogen fluoride, using the ML model trained exclusively on 100 water geometries. 
As before, Feature Set C is used to avoid overfitting. 
For nine of the 100 \ch{HF} geometries, one pair of occupied LMOs energetically reorders in a way that is not accounted for the feature sorting protocol described in  Sec.~\ref{sec:feature_design}; to address this, the $i,j$ sorting of one pair of LMOs  was done  manually  in these 9 \ch{HF} geometries.

The results in Fig.~\ref{figure:water_to_other} clearly indicate that the CCSD energies for the \ch{NH3}, \ch{CH4}, and \ch{HF} molecules are accurately predicted by the ML model on the basis of training data that comes entirely from \ch{H2O}. %
The Mean Errors fall within 0.5 mH, and Rel.~Mean Errors remain below 0.24\% in all cases. The Pearson coefficient exceeds 0.94 in all cases, indicating excellent correlation although the results are somewhat skewed.
These results demonstrate that the ML model successfully transfers information about the fundamental components of the electronic structure of water -- i.e., lone pairs and sigma bonds -- for the prediction of similar components in other molecules, even when those molecules are composed of different elements.

\section{Discussion and Conclusions}
We have introduced a ML method  for predicting correlated electronic  structure  energies using input from an SCF
calculation. With features formulated in terms of molecular orbitals %
-- rather than atom-type or element specific features -- the method is designed with the aim of providing a compact feature set for learning and good transferability across chemical systems.
A previous effort in this direction focused on predicting accurate non-covalent interactions using interaction energies from lower levels of electronic structure;\cite{mcgibbon2017improving} our method seeks to predict correlated interactions between pairs of occupied MOs rather than between pairs of molecules. 

  The transferability of the method has been demonstrated in several examples, illustrating that it can be used for accurate MP2 and CCSD energy predictions  for molecules with different bonding connectivities and different chemical elements than those included in the training set. Of the various applications of the ML method in this work, the relative mean error is at most 0.24\% of the CCSD correlation energy and the Pearson correlation coefficients are consistently greater than 0.9 
-- with the notable exception of the case of butane and isobutane where the dynamic range of the true correlation energy is small. %
Indeed, the range of ML prediction errors is found to be relatively independent of the range of true correlation energies for the systems considered here.
Furthermore, the method is shown to work equally well for the prediction of both MP2 and CCSD correlation energies, suggesting that it will be similarly effective in the prediction of other single-reference correlated electronic structure methods.
The description of the ML features in terms of localized molecular orbitals was found necessary to provide a modular and thus transferable ML model. %

In terms of compactness of the ML feature set, all calculations presented here employ between 11 and 26 unique features.  Alternatively stated, %
we find that  at most 26 matrix elements from a Hartree-Fock calculation are needed to predict the contribution to the correlation energy from any pair of occupied valence orbitals; the training data includes no meta-data about atom-types, bond-types, geometry, or about the chemical environment in which the orbital pair resides.

Although several avenues for development and application of the method are possible, natural objectives for future work include 
(i) reduction of the baseline self-training errors of the %
simple Gaussian process regression method employed here;
(ii) formulation of the ML method in terms of input from smaller basis sets or from low-cost SCF theories, such as density functional tight-binding;
 and (iii) implementation of a gradient theory that employs coupled perturbed SCF and localization, as for the gradients of local correlation methods.\cite{Schutz2004} %

\section*{Acknowledgements}
This work was supported by AFOSR award no. FA9550-17-1-0102.
The authors additionally acknowledge support from  the Resnick Sustainability Institute postdoctoral fellowship (MW), a Caltech Chemistry graduate fellowship (LC), and the Camille Dreyfus Teacher-Scholar Award (TFM).

\section*{Supporting information}
Supporting information is provided and includes 
expanded details for small molecule predictions (corresponding to Tab. \ref{table:self-prediction}), 
MP2 results corresponding to all CCSD results presented in the main text, 
and plots of ML prediction error versus total CCSD energy corresponding to Figs. \ref{figure:water_clusters}-\ref{figure:water_to_other}.

\bibliography{main}

\end{document}